\documentclass[
    ,final            
  ]
  {aipproc}

\layoutstyle{8x11single}

\newcommand{\be}{\begin{equation}}
\newcommand{\ee}{\end{equation}}
\newcommand{\ba}{\begin{eqnarray}}
\newcommand{\ea}{\end{eqnarray}}

\newcommand{\np}{{\bf      p}}
\newcommand{\nq}{{\bf      q}}

\begin{document}

\title{Neutrino Interactions Importance to Nuclear Physics}

\classification{25.30.Pt,24.10.-i,25.30.Fj} 
\keywords      {Neutrino induced nuclear reactions}

\author{J.E. Amaro}{
  address={Departamento de Fisica Atomica, Molecular y Nuclear, 
Universidad de Granada, 18071 Granada, Spain}
}
\author{C. Maieron}{
  address={Departamento de Fisica Atomica, Molecular y Nuclear, 
Universidad de Granada, 18071 Granada, Spain}
}

\author{M. Valverde}{
  address={Research Center for Nuclear Physics, 
Osaka University, Ibakari 567-0047, Japan}
}

\author{J. Nieves}{
  address={Instituto de Fisica Corpuscular, Centro Mixto CSIC-Universidad de Valencia, Institutos de Investigacion de Paterna, Aptd. 22085, 46071 Valencia, Spain}
}

\author{M.B. Barbaro}{
  address={Dipartimento di Fisica Teorica, University of Turin, and INFN Sezione di Torino, 10125 Turin, Italy.}
}

\author{J.A.~Caballero}{
  address={Departamento de Fisica Atomica, Molecular y Nuclear, Universidad de Sevilla. Apdo. 1065, 41080 Sevilla, Spain}
}

\author{T.W. Donnelly}{
  address={Center for Theoretical Physics, Laboratory for Nuclear Science and Department of Physics, Massachusetts Institute of Technology, Cambridge, MA 02139, USA}
}

\author{J.M. Udias}{
   address={Departamento de F\'{\i}sica At\'omica, Molecular y Nuclear, 
Universidad Complutense de Madrid, 28040, Madrid, Spain}
}

\begin{abstract}
We review the general interplay between Nuclear Physics and
neutrino-nucleus cross sections at intermediate and high energies.
The effects of different reaction mechanisms
over the neutrino observables are illustrated with examples 
in calculations using several nuclear models and ingredients.
\end{abstract}

\maketitle

As motivation for the more specific 
workshop sessions, in this talk we introduce the general
formalism of neutrino scattering from nuclei and define the
observables of interest for nuclear physics at the energy
regime of interest. 
Using different nuclear models 
we present with examples the 
 theoretical ingredients of relevance for neutrino reactions:
long Range nuclear correlations (RPA), final
state interactions (FSI), finite-size effects, Coulomb corrections,
and relativistic effects. Theoretical results will be shown for  
charge-changing quasielastic neutrino scattering.
We skip here
the discussion on neutral current scattering, Delta excitation and
coherent pion production, which will be summarized in the M.B. Barbaro
and E. Hernandez papers in these proceedings.
We present results for kinematics going from low to high energy and for
different kind of observables: response functions, inclusive cross sections,
integrated cross sections, angular distributions, polarization
observables, etc.  Nuclear models for which we will show results are
Local Fermi Gas (LFG), Relativistic Fermi Gas (RFG), Semi-relativistic 
Shell Model (SR-SM),
and Super-Scaling Analysis (SuSA) model.
We also skip the discussion on the Relativistic Mean Field, which will be 
summarized in the talk by J.M. Udias in these same proceedings.
Some particular topics that we briefly discuss are theoretical
uncertainties on the ratios of interest for experiments on atmospheric
neutrinos, nuclear effects on lepton polarization,
and reconstruction of neutrino cross section from electron scattering data.

\subsection{Formalism}

We consider as example the  
CC neutrino reaction $\nu_l + A \rightarrow l^- + B$,
for initial $K^\mu$ and final $K'{}^{\mu}$ lepton momenta,
with energies $\epsilon$ and $\epsilon'$, respectively.
We introduce also the momentum transfer
$Q^\mu= K^\mu-K'{}^{\mu}=P'{}^\mu-P^{\mu}$.
It is convenient to define the following adimensional variables
$\lambda=\frac{\omega}{2m_N}$,
$\kappa=\frac{q}{2m_N}$, and 
$\tau = \kappa^2-\lambda^2$. 

The inclusive cross section for the inclusive $(\nu_l,l^-)$ reaction 
where only the final lepton is detected, can be written as
\[
\frac{d\sigma}{d\Omega'd\epsilon'}=
\frac{G^2\cos^2\theta_c}{2\pi^2}
k'\epsilon'\cos^2\frac{\tilde{\theta}}{2}
{\cal F}_+^2 \ 
\]
Where the coupling constant $G=1.1664\times 10^{-5} \rm GeV^{-2}$,
the Cabibbo angle $\theta_c=0.974$, 
and the generalized scattering angle is defined by
$
\tan^2\frac{\tilde{\theta}}{2}=\frac{|Q^2|}{(\epsilon+\epsilon')^2-q^2}.
$
The interesting nuclear information is contained 
in the structure function 
\[
{\cal F}_+^2=
\widehat{V}_{CC} R_{CC}+
2\widehat{V}_{CL} R_{CL}+
\widehat{V}_{LL} R_{LL}+
\widehat{V}_{T} R_{T}+
2\widehat{V}_{T'} R_{T'}.
\]
The kinematic factors $\widehat{V}_K$, coming from the leptonic tensor, are
defined by \cite{Ama05b}
\begin{eqnarray*}
&&
\widehat{V}_{CC}
=
1-\delta^2\tan^2\frac{\tilde{\theta}}{2},
\kern 1cm
\widehat{V}_{LL}
=
\frac{\omega^2}{q^2}+
\left(1+\frac{2\omega}{q\rho'}+\rho\delta^2\right)\delta^2
\tan^2\frac{\tilde{\theta}}{2},
\kern 1cm
\widehat{V}_{CL}
=
\frac{\omega}{q}+\frac{\delta^2}{\rho'}\tan^2\frac{\tilde{\theta}}{2}
\\
&&
\widehat{V}_{T}
=
\tan^2\frac{\tilde{\theta}}{2}+\frac{\rho}{2}-
\frac{\delta^2}{\rho'}
\left(\frac{\omega}{q}+\frac12\rho\rho'\delta^2\right)
\tan^2\frac{\tilde{\theta}}{2},
\kern 1cm
\widehat{V}_{T'}
=
\frac{1}{\rho'}
\left(1-\frac{\omega\rho'}{q}\delta^2\right)
\tan^2\frac{\tilde{\theta}}{2}
\end{eqnarray*}
where the following adimensional variables are introduced:
$
\delta = \frac{m'}{\sqrt{|Q^2|}},
$
$\rho = \frac{|Q^2|}{q^2}$,
and $\rho' = \frac{q}{\epsilon+\epsilon'}$.

Finally the following five nuclear weak response functions appear
\begin{eqnarray*}
R_{CC}=W^{00},
& R_{LL}= W^{33},
& R_{CL}= -\frac12\left( W^{03}+W^{30} \right)\\
R_T =  W^{11}+W^{22} &&
R_{T'} = -\frac{i}{2}\left( W^{12}-W^{21} \right)
\end{eqnarray*}
as linear combinations of the weak CC hadronic tensor.
\[
W^{\mu\nu}=\overline{\sum_{fi}}\delta(E_f-E_i-\omega)
\langle f | J^\mu(\nq) | i\rangle^*
\langle f | J^\nu(\nq) | i\rangle
\]
where the 
matrix elements of the hadronic current are taken between the 
initial and final hadronic states in the nuclear transition
 $|i\rangle \rightarrow |f\rangle$.  
The hadronic current 
for a single nucleon is of the form $V-A$
\begin{equation}
\hat{J}_\mu=\overline{u}_p(\np')
\left[F_1(Q^2)\gamma_\mu+F_2(Q^2)i\sigma_{\mu\nu}\frac{Q^\nu}{2m_N}
     -G_A(Q^2)\gamma_\mu\gamma_5-G_P(Q^2)\frac{Q_\mu}{2m_N}\gamma_5
\right]
u_n(\np)
\end{equation}

\subsection{The Relativistic Fermi Gas (RFG)}

As example of the above general formalism in a many-nucleon system, 
we present here the nuclear response functions in the RFG, where the 
initial and final nucleons are described as (relativistic) plane waves.  
For the inclusive $(\nu_l,l^-)$ reaction we can write 
the five response functions as 
\begin{equation}\label{rfg}
R_K = N \Lambda_0 U_K  f_{RFG}(\psi)= G_K f_{RFG}(\psi),  \quad K=CC,CL,LL,T,T',
\end{equation}
Here $N$ is the neutron number,
and we have defined $\Lambda_0 = \frac{\xi_F}{m_N \eta_F^3 \kappa}$,
with $\eta_F=k_F/m_N$ is the Fermi momentum in units of the nucleon mass,
and  $\xi_F=\sqrt{1+\eta_F^2}-1$ is the Fermi kinetic energy measured 
in the same units
All the response functions are proportional to the 
RFG Scaling function
$f_{RFG}(\psi)=\frac34 (1-\psi^2)\theta(1-\psi^2)$,
that depends only on the {\em scaling variable}
\begin{equation}\label{psi}
\psi=\frac{1}{\sqrt{\xi_F}}
\frac{\lambda-\tau}{\sqrt{(1+\lambda)\tau+\kappa\sqrt{\tau(1+\tau)}}}
\end{equation}
Explicit expressions can be obtained for the single-nucleon responses $U_K$
\cite{Ama05b}.

\subsection{The Local Fermi Gas (LFG)}

As an extension of the constant-density Fermi gas, the LFG is an 
easy model to include non-trivial nuclear effects \cite{Nie04}.
It is based on the local density Approximation (LDA), with a
local Fermi momentum is  $k_F(r)=(3\pi^2 \rho(r))^{1/3}$.
The responses are averaged 
over the nuclear interior, weighted by the nucleon density $\rho(r)$. 
In this model one can easily include 
relevant nuclear effects such as the correct energy balance 
RPA nuclear correlations, Coulomb distortion, and
FSI effects.

While the LFG correctly takes into account Pauli-blocking effects, 
it gives a wrong energy balance for the nuclear excitations. 
The energy balance can be corrected by the minimum nuclear excitation energy 
gap $ Q= M(X_f)-M(X_i)$,
instead of the usual LFG value $Q_{LFG}(r)= E_F^p(r)-E_F^n(r)$.  
That is equivalent to replacing 
$\omega \longrightarrow \omega -[Q-Q_{LFG}(r)]$.

The RPA series of Fig. 1 is solved by using a  
ph-ph interaction of Landau-Migdal,
with parameters fitted to electromagnetic nuclear properties and transitions
\cite{Spe80}.
The RPA series can be generalized by including 
$\Delta$ excitations in the medium, 
and ph-$\Delta$h,  $\Delta$-h$\Delta$h 
effective interactions. The sum of the RPA series is equivalent to
a renormalization of the axial and vector parts of the weak hadronic tensor
in the medium

Coulomb corrections can be included in the LFG 
by introducing  the Coulomb self-energy of the final lepton
$\Sigma_{C}=2\epsilon' V_C(r)$ 
(where $V_C(r)$ is the nuclear Coulomb potential)
in the charged 
lepton propagator,
$\frac{1}{k^2-m_l^2-2k_0 V_C(r)+i\epsilon}$.
A new local energy-momentum relation for the final lepton is then obtained,
that is used in the LFG calculation.
This procedure 
is equivalent to the modified effective momentum approximation.

LFG results for $^{12}$C$(\nu_\mu,\mu^-)$ and $^{12}$C$(\nu_e,e^-)$
are presented in Fig 1, and are compared with experimental data 
\cite{Aue02,Ath97} in
Table 1.  The LFG with correct energy balance (Pauli+Q),
over-estimates the data, while a good agreement is achieved once RPA
and Coulomb corrections are included.

\begin{figure}
\includegraphics[scale=0.6]{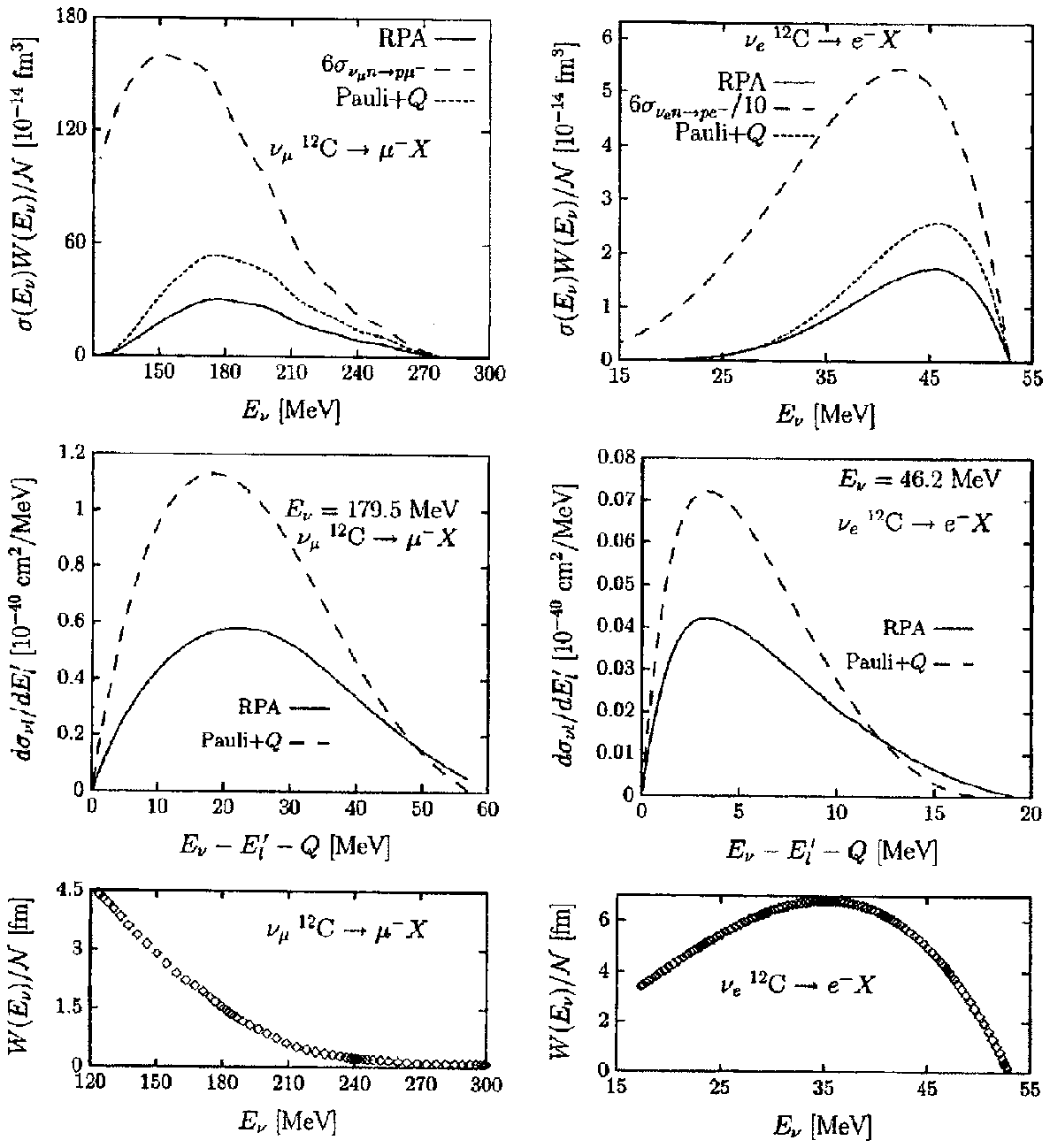}
{\includegraphics[scale=0.9]{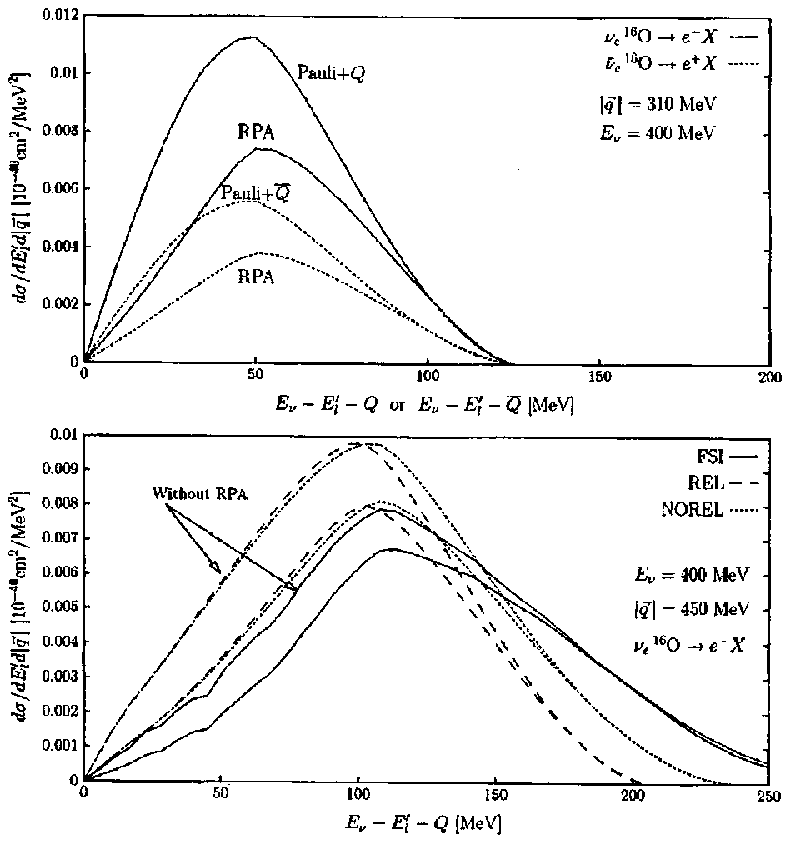}}%
\caption{
Left: LFG predictions for neutrino reactions from $^{12}$C.  Left
 panels: $(\nu_\mu,\mu^-)$. Right panels: $(e,\nu_e)$.  Pauli+Q are
 the LFG results with correct energy balance.  RPA are the results
 with RPA and Coulomb corrections.  Top panels: total cross sections
 multiplied by the neutrino fluxes of the bottom.  Middle panels:
 differential cross section for fixed neutrino energy.
Right top: $\nu_e$ and
$\overline{\nu}_e$ differential cross section for fixed momentum
transfer; right bottom: effect of the FSI over
the neutrino cross section, with and without RPA.  }
\end{figure}

\begin{table}
\begin{tabular}{cccc} \hline
                  & Pauli+Q & RPA  & LSND \\ \hline
$(\nu_\mu,\mu^-)$ & 20.7    & 11.9 & $10.6\pm 0.3\pm 1.8$ \\
$(e,\nu_e)$       & 0.19    & 0.14 & $0.15\pm0.01\pm0.01$\\
\hline
\end{tabular}
\caption{
Flux averaged neutrino cross section from $^{12}$C in $10^{-40}$ cm$^2$.
The experimental data are from  LSND \protect\cite{Aue02,Ath97}.}
\end{table}

Final State Interaction (FSI) can be taken into account in the LFG
model by using a renormalized nucleon propagator in the medium
$G_{FSI}(p)=\frac{1}{p^0-E(\vec{p})-\Sigma(p)}$, where $\Sigma(p)$ is
the nucleon self-energy in the medium.  A good approximations for
intermediate energies is to take ${\rm Im}\Sigma_h\simeq 0$ for hole
states.  The nucleon self energy can be computed by diagrammatic
techniques \cite{Fer92}. As shown in Fig. 1, the FSI importantly
changes the shape of the differential cross section. 
The main effect is is an enhancement of the high energy tail and
a reduction of the cross section
at the peak region. In the same figure we also show that the 
RPA corrections are less important in presence of FSI.

Theoretical uncertainties in the LFG model have been computed assuming
central values and errors of the model input parameters, and including
10\% uncertainties in both the real part of the nucleon self-energy and
densities \cite{Val06}.  A Montecarlo simulation is then performed by
generating sets of input parameters using Gaussian distributions.
After computing the different observables, one obtains the
distribution of the observable values, and the theoretical errors are
identified by discarding the highest and lowest 16\% of the obtained
values, keeping a 68\% confidence level interval.  Uncertainties on
the integrated cross sections are of the order of 10-15\%, which turn
out to be similar to those assumed for the input parameters.  As shown
in Fig. 3, theoretical errors cancel partially out in the ratio
$
\frac{\sigma(\mu)}{\sigma(e)}
=\frac{\sigma(\nu_\mu,\mu)}{\sigma(\nu_e,e)}
$
of interest for experiments on atmospheric neutrinos. 

\begin{figure}
{\includegraphics[scale=0.7, bb=120 550 500 800, clip]{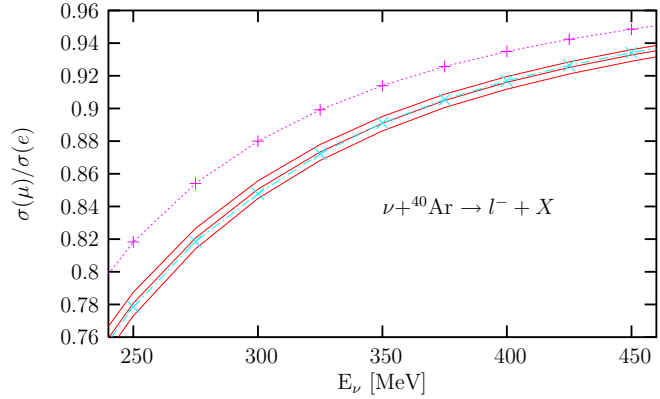}}%
\caption{Ratio of inclusive cross sections
$\frac{\sigma(\mu)}{\sigma(e)}$ for argon. The 68\% confidence level
band is displayed for the full LFG model including RPA, Coulomb and
FSI. We also show results with the bare Fermi gas model without nuclear
corrections.  }
\end{figure}

\begin{figure}
{\includegraphics[scale=0.6, bb=75 400 500 800]{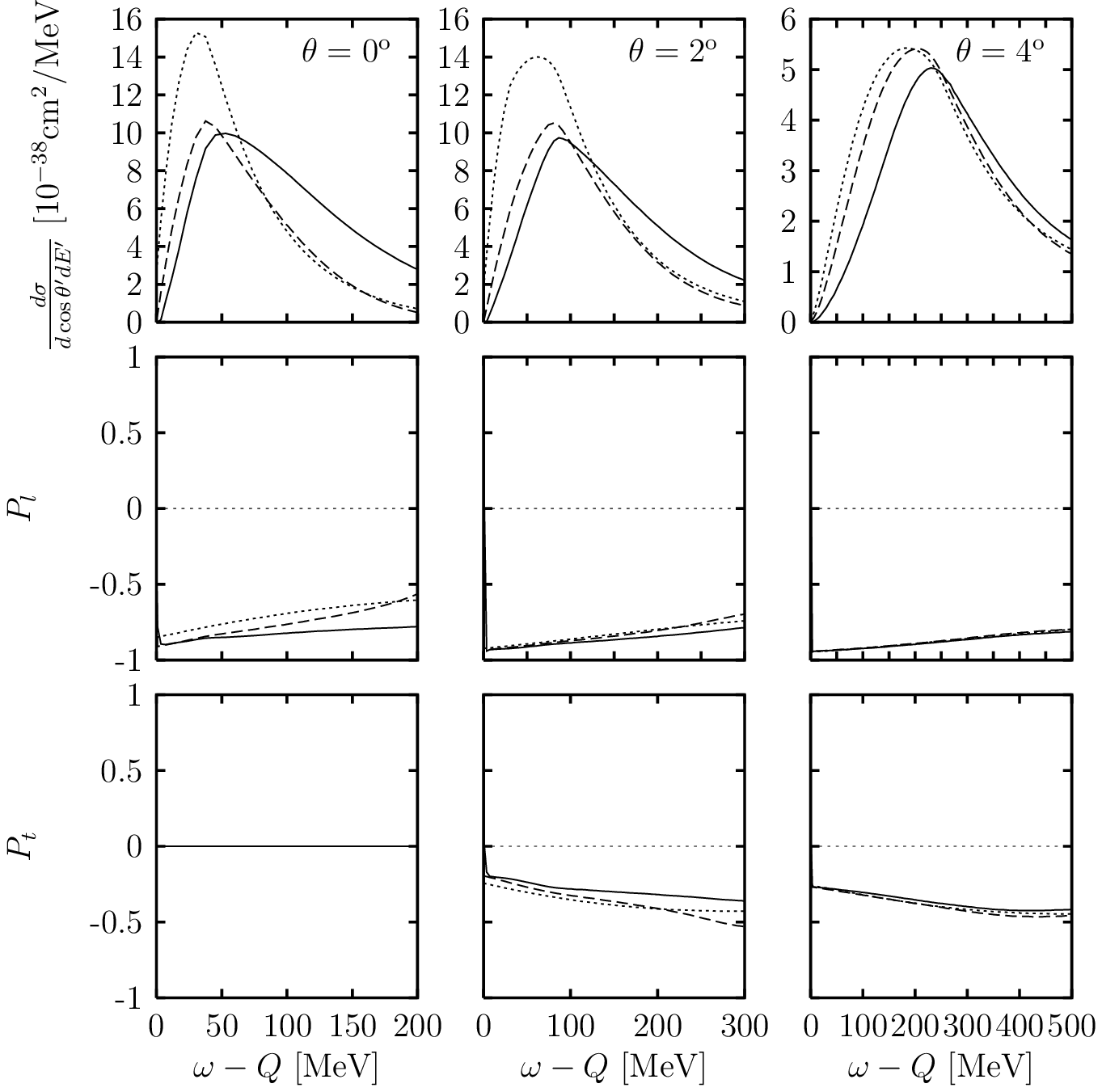}}%
{\includegraphics[scale=0.6, bb=75 500 500 800]{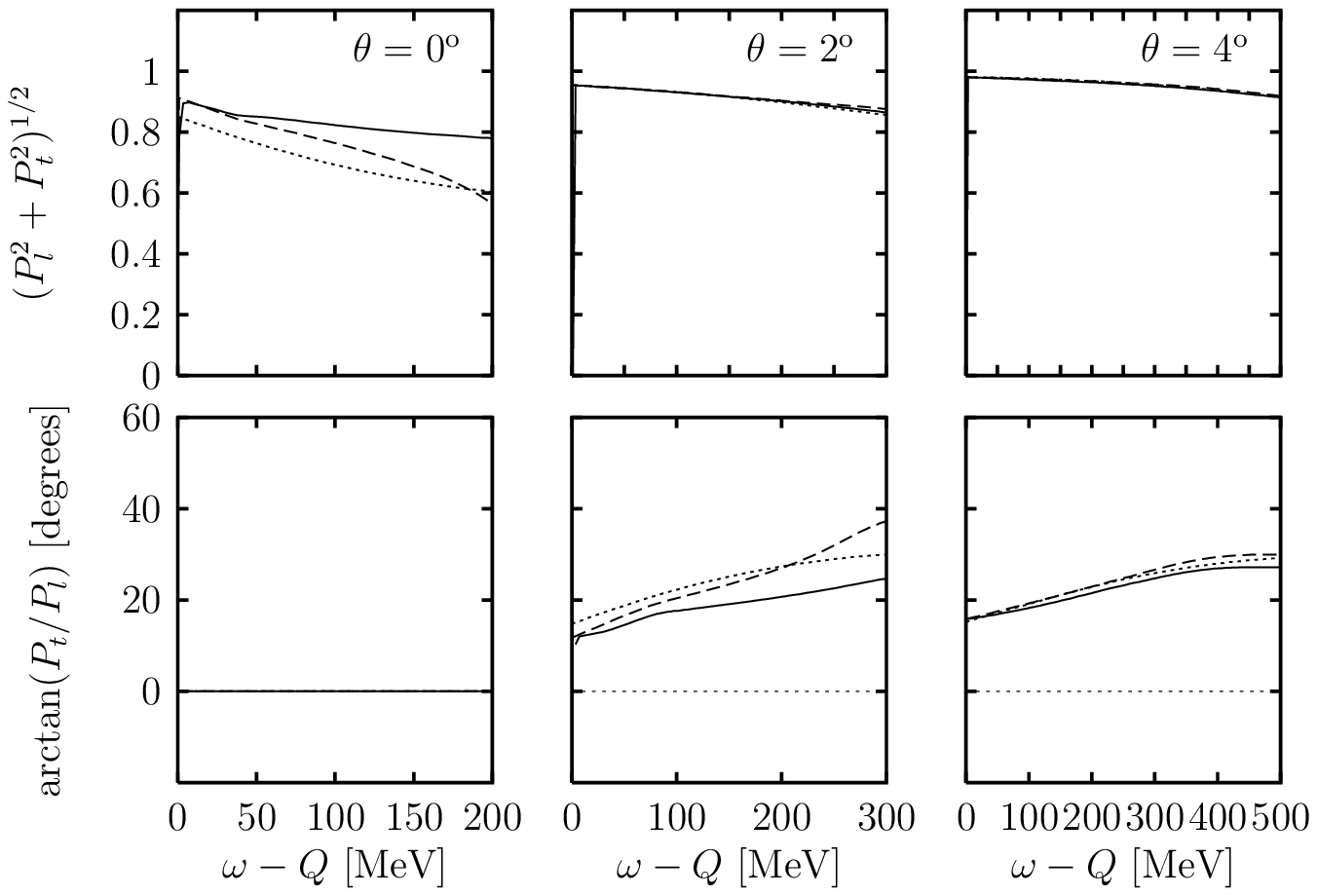}}%
\caption{Polarization observables 
in the LFG (dotted), adding RPA correlations 
(dashed), and adding FSI (solid).
}
\end{figure}

\subsection{Polarization observables}

The study of final $\tau$ polarization in $(\nu_\tau,\tau)$ reactions
\cite{Val06b} is of interest for $\nu_\mu\longrightarrow\nu_\tau$ oscillation
experiments.  The $\tau$ decay particle distribution depend on the
$\tau$ spin direction, and thus theoretical information on the $\tau$
polarization will be valuable \cite{Mig03}.  Information about $\tau$
polarization is also needed in $\nu_\mu\longrightarrow\nu_e$
oscillation experiments to disentangle $(\nu_e,e)$ events from
background electron productions following the
$\nu_\mu\longrightarrow\nu_\tau$ oscillation \cite{Hag03}.

The polarized differential cross section  
in $(\nu_l,\vec{l})$ reactions when the 
final lepton polarization is measured in the direction $\vec{s}$
can be written as
\[
\Sigma(\vec{s})\equiv 
\frac{d^2\sigma}{d\Omega'dE^\prime_l}
=\frac12\Sigma_0\left(1+s_\mu P^\mu\right)
\]
where $\Sigma_0$ is the unpolarized cross section. The lepton
polarization vector components $P_l$ (longitudinal component, in the
direction of the final lepton), and $P_t$ (transverse component in the
scattering plane) can be obtained as asymmetries
\[
s_\mu P_\mu 
= \frac{\Sigma(\vec{s})-\Sigma(\vec{-s})}{\Sigma(\vec{s})+\Sigma(\vec{-s})}.
\]
Results for the polarization observables with the LFG model are shown
in Fig. 4, with the differential cross section, the two
$\tau$ polarization components, and the total polarization and angle.
The nuclear RPA and FSI effects are of small importance at the quasielastic 
peak region because of
partial cancellation when computing the asymmetries.
Their importance increases in the tail of the cross section.

\subsection{Super-Scaling Analysis (SuSA)}

In the RFG, Eqs. (\ref{rfg},\ref{psi}), all the response functions can
be factorized in a single-nucleon function times the superscaling
function $f_{RFG}$, which only depends on the scaling variable $\psi$
and is the same for all nuclei. We then say that the RFG superscales,
namely $f_{RFG}$ does not explicitly depends upon the momentum
transfer $q$ (scaling of the first kind) nor the Fermi momentum $k_F$
(scaling of the second kind). Moreover the scaling function is the
same for electron and neutrino scattering, hence the neutrino and
electron cross sections are related just by a single-nucleon factor.

These ideas have been extended to extract an ``experimental''
 scaling function from $(e,e')$ data by computing the ratio
\[f_{exp}(\psi')=
\frac{\displaystyle
\left(\frac{d\sigma}{d\Omega'd\epsilon'}\right)_{exp}}{\sigma_{Mott}(v_LG_L+v_TG_T)}
\]
where $\psi'$ is the scaling variable shifted to account for an energy binding
parameter $E_s$. 
An extensive analysis of $(e,e')$ data has been performed in the
quasielastic peak \cite{Mai02}. The parameters $k_F$ y $E_s$ are
fitted to the data to minimize the differences between scaling
functions for different kinematics and nuclei. When this analysis is performed
on the experimentally available 
longitudinal,
$f_L = \frac{R_L}{G_L}$,
 and transverse, 
$f_T = \frac{R_T}{G_T}$, response functions,
one finds that the L response function superscales quite well, while
scaling is broken in the T response above the peak, due to
non-quasielastic processes (pion production, $\Delta$ excitation, MEC,
etc). The experimental scaling function is shown in Fig. \ref{deb-scaling} below

\begin{figure}
\includegraphics[scale=0.6,  bb= 150 360 400 790]{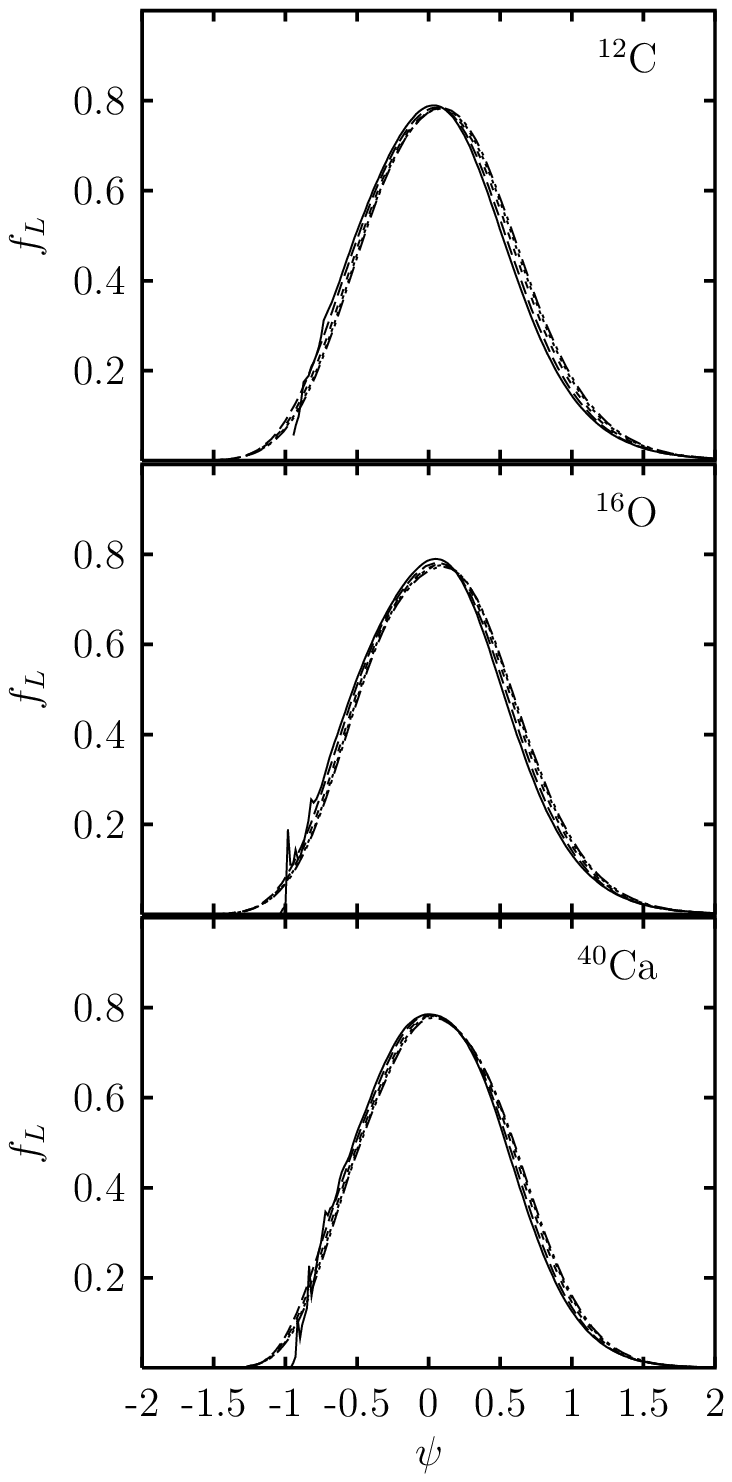}
\includegraphics[scale=0.57,  bb= 180 310 400 790]{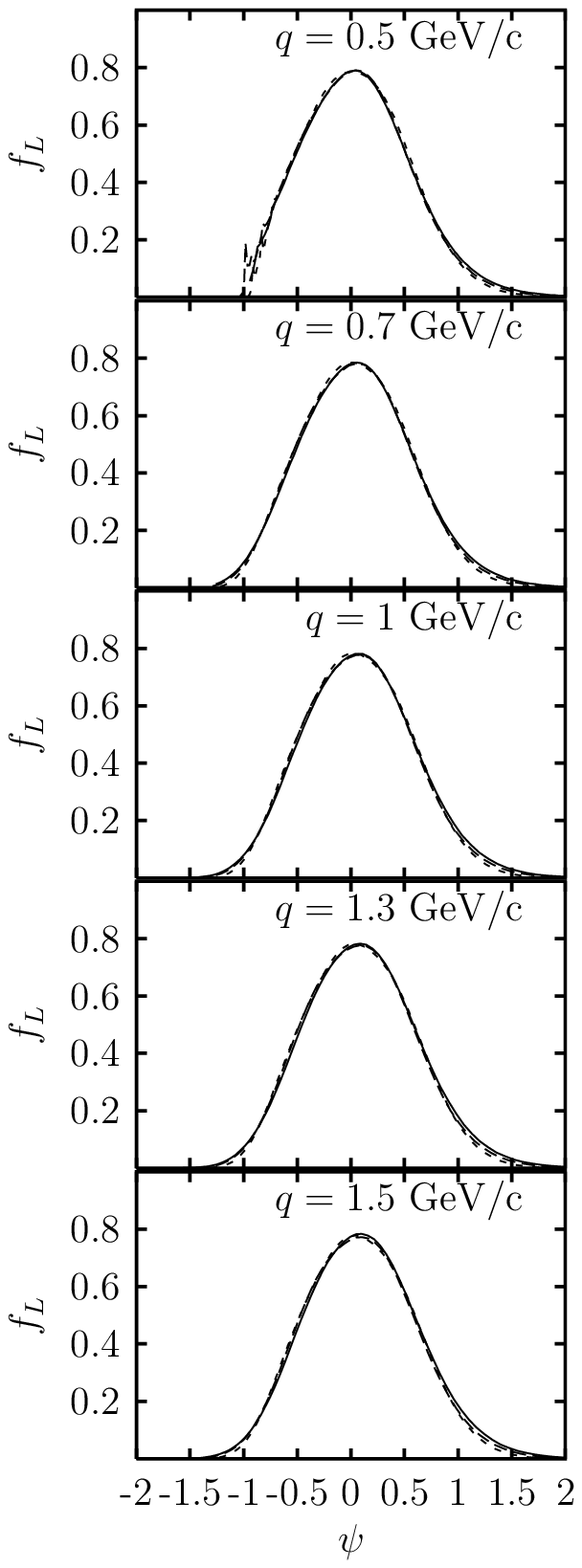}
\caption{\label{scaling-sr}
Scaling properties of the SR shell model.
Left:
scaling of the first kind.
Curves for 
$q=0.5,0.7,1,1.3,1.5$ GeV 
collapse into one.
Right: scaling of the second kind.
Curves for 
$^{12}$C, $^{16}$O and $^{40}$Ca 
collapse into one
}
\end{figure}

Having succeeded in representing the $(e,e')$ quasielastic data by
means of an universal superscaling function, the SuSA procedure can be
reversed and give predictions for neutrino reactions \cite{Ama05}.  In fact,
starting with the experimental $(e,e')$ scaling function one can use
the RFG equations (\ref{rfg}) to compute the $(\nu_l,l^-)$ response
functions with the substitution $f_{RFG}(\psi) \longrightarrow
f_{exp}(\psi)$.  The assumption underlying this procedure is that the
superscaling function associated to the 5 weak responses is the same
and is equal to the longitudinal electromagnetic response function.
Calculations aiming to  to justify theoretically the validity of SuSA
are presented in the next section.

\subsection{The semirelativistic shell model (SRSM) }

In order to study the scaling properties in realistic models one needs
to go beyond the simplicity of the Fermi gas, where scaling holds by
construction.  In the continuum shell model (CSM), i.e., nucleons in a
mean field, distortion of the ejected nucleon is present and a general
proof of scaling cannot be provided. Thus, at least within the context
of the CSM, we shall be able to check the consistency of the SuSA
approach and quantify the degree to which scale breaking effects are
expected to enter.

In the CSM
the initial state $|i\rangle$ is described by a  Slater determinant
with all shells occupied, while 
in the impulse approximation the final states
are particle-hole excitations coupled to total angular
momentum 
$|f\rangle =|(ph^{-1})J\rangle$.
The single hole wave function is then written as  
$|h\rangle=|\epsilon_hl_hj_h\rangle$ 
while the
single particle  wave function  
$|p\rangle=|\epsilon_pl_pj_p\rangle$.
The radial functions are
obtained by solving the Schr\"odinger equation
with a Woods-Saxon potential.

At the ongoing and next generation neutrino experiments the
neutrino beam energies increase to the GeV level and, typically, large
energies and momenta are transferred to the nucleus.  For these
kinematics, relativity is important. The relevant relativistic
corrections can be easily implemented in the CSM through the
semi-relativistic approach (SR) \cite{Ama05b}.
 It is based on a new expansion 
of the relativistic single-nucleon current
$j^{\mu}(\vec{p'},\vec{p}) = \overline{u}(\vec{p'})\Gamma^\mu(Q)u(\vec{p})$ 
in powers of the initial nucleon momentum,
$\vec{\eta}= \vec{p}/m_N$,
to first order $O(\eta)$.
We do not expand in $\vec{p'}/m_N$, hence
$q$ and $\omega$ can be arbitrarily large.
Second one must use relativistic kinematics.
The energy transfer in the CSM is the difference between
the (non-relativistic) single-particle energies of particle and hole
$\omega=\epsilon_p-\epsilon_h$. 
The relativistic kinematics are taken into
account by the substitution 
$\epsilon_p\rightarrow
\epsilon_p(1+\epsilon_p/2m_N)$
as the eigenvalue of the Schr\"odinger equation
for the particle in the continuum.

A test of the SR approach was performed in \cite{Ama05b} 
by comparing the exact RFG
results with those of the SR Fermi gas model, where the above SR
current and relativistic kinematics have been implemented.

An study of superscaling properties of the SRSM, for $q> 0.5$ GeV/c,
is performed in Fig. \ref{scaling-sr}. Scaling of both first and
second kind are achieved in the shell model: all the curves
collapse into one with small deviations for $q=0.5$ GeV.  Since both
kind of scaling are found, we conclude that superscaling occurs within
our model.

\begin{figure}
\includegraphics[scale=0.6,  bb= 70 360 490 790]{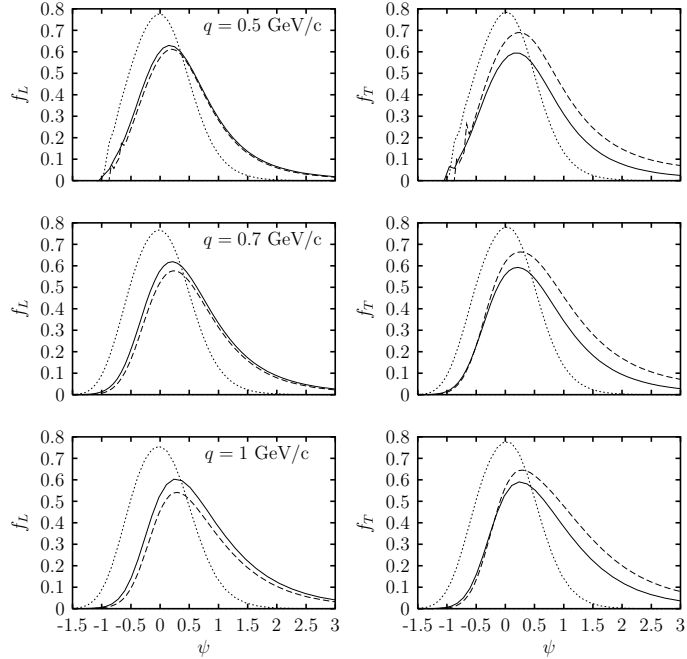}
\caption{\label{deb-rmf}
Scaling functions computed in the 
SR shell model compared with the relativistic mean field model (RMF).
Dotted: Woods-Saxon potential.
Solid: DEB+D potential. 
Dashed: RMF. 
}
\end{figure}

The SRSM with Woods-Saxon potential fails to reproduce the
experimental scaling function, giving results similar to the RFG, with
the exception of small tails for low and high energy transfer.  The next
step \cite{Ama07} is to improve the relativistic description of the ejected
nucleon. The FSI is modified by using a Dirac-equation based potential
plus Darwin term (DEB+D) in the final state, instead of the usual
Woods-Saxon potential. The trick is to
rewrite the Dirac equation as a second-order equation for the upper component
$\psi_{up}(\vec{r})$.
The Darwin term is then defined by: 
$\psi_{up}(\vec{r})=K(r,E)\phi(\vec{r})$,
where the function $\phi(\vec{r})$ verifies the Schr\"odinger equation
\[ \left[-\frac{1}{2m_N}\nabla^2+U_{DEB}(r,E)\right]\phi(\vec{r})=
\frac{E^2-m_N^2}{2m_N}\phi(\vec{r}).
\]
Both the DEB potential  $U_{DEB}(r,E)$ and Darwin term $K(r,E)$
are energy-dependent functions.
The scaling functions computed using the DEB+D potential are compared to the exact RMF results in Fig. \ref{deb-rmf}. Both models give similar results,
and clearly different to the SR model based on a Woods-Saxon potential.  
 
\begin{figure}
\includegraphics[scale=0.6,  bb= 70 360 490 790]{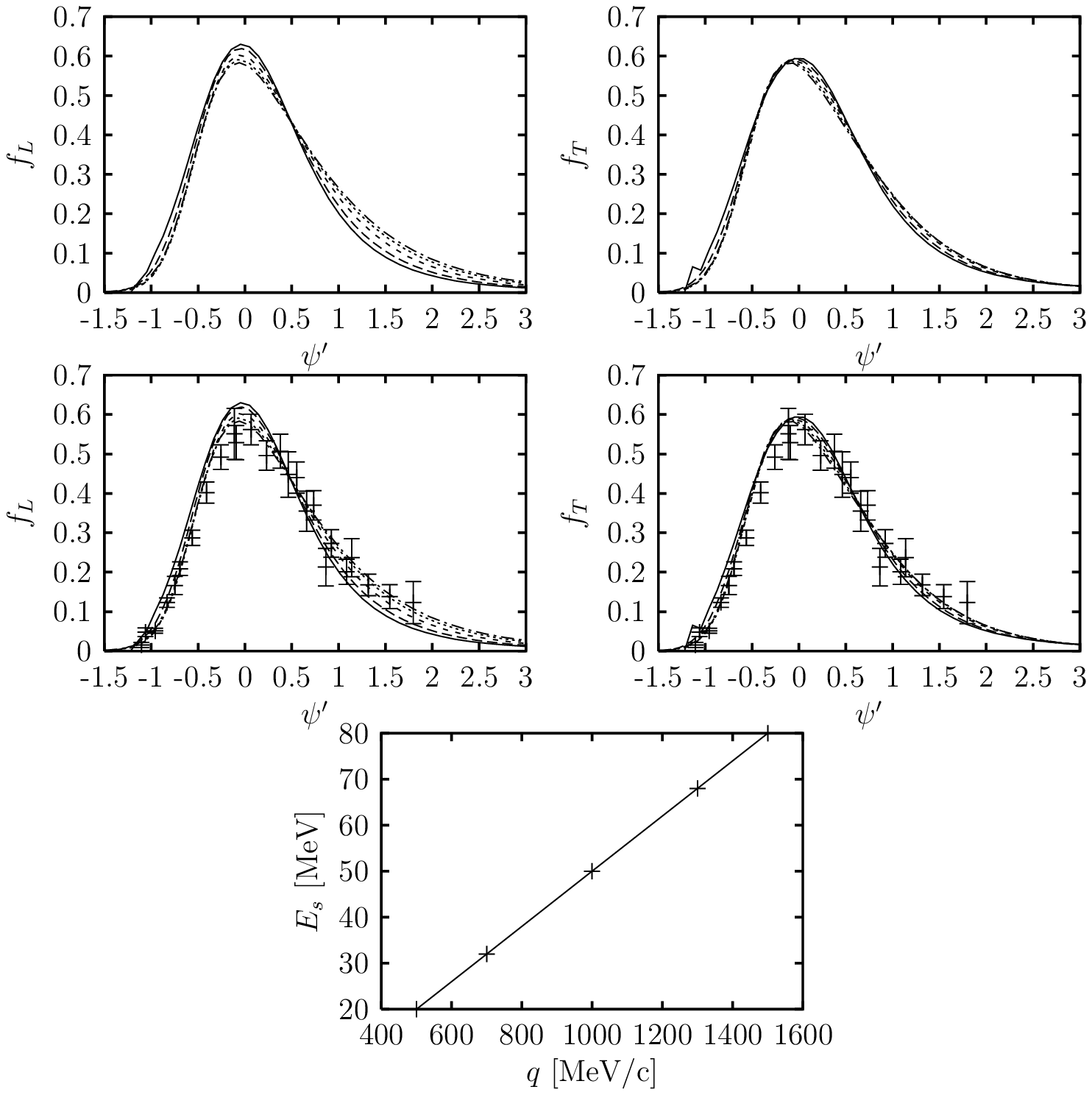}
\includegraphics[scale=0.4,  bb= 40 90 490 780]{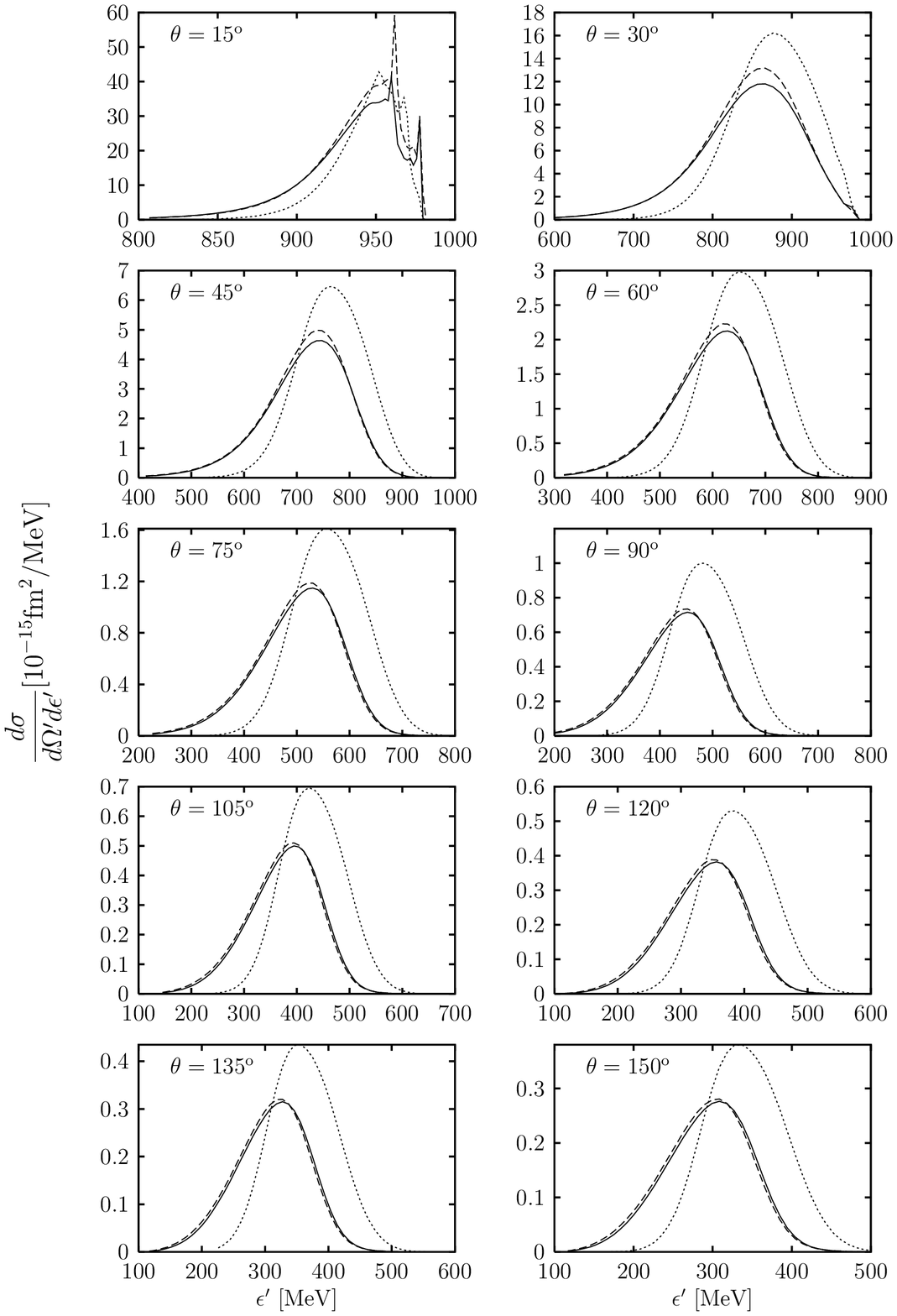}
\caption{\label{deb-scaling}
Left: Scaling 
of 1st kind with
DEB+D potential compared with experimental data.
$q=0.5$, 0.7
1.0, 1.3 
and 1.5 GEV/c.
Right:
Test of SuSA in the SR shell model for the $^{12}$C$(\nu_\mu,\mu^-)$
reaction with neutrino energy $\epsilon=1$ GeV.
Dotted: Woods-Saxon potential.
Solid: DEB+D potential.
Dashed: SuSA reconstruction from the computed $(e,e')$ scaling function.
}
\end{figure}
Scaling of the first kind of the SRSM with the DEB+D potential is shown in
Fig. \ref{deb-scaling}, where we also compare with the experimental
data for $f_L$.  The scaling is surprisingly good, taking into account
that the momentum transfer ranges from 0.5 to 1.5 GeV.  A q-dependent
energy shift $E_s(q)$ is needed to bring all the curves to
collapse. The $q$ dependence of $E_s(q)$ is linear.

A test of the SuSA reconstruction of the $(\nu_\mu,\mu^-)$ 
cross section from the 
$(e,e')$ one is shown in Fig. \ref{deb-scaling}.
There the neutrino cross section obtained by direct computation
is compared with the SuSA reconstruction 
from the computed $(e,e')$ longitudinal scaling function.
Both results are quite similar for scattering angle above 30$^{\rm o}$,
corresponding to intermediate to large momentum transfer. For very low 
scattering angle there are small differences
between both procedures due to the small momentum transfers involved.

\subsection{Concluding remarks}

Neutrino interactions importance for nuclear physics has been
illustrated with examples from a selection of neutrino reactions on
nuclei using different approaches. The connection of electron and
neutrino cross sections has also been analyzed within the superscaling
approach.  Since neutrino cross sections incorporate a richer
information on nuclear structure and interactions than electron cross
sections, the availability of neutrino-nucleus observables of
different kinds will be valuable for the development of more precise
nuclear models and nuclear interaction theories.

\begin{theacknowledgments}
This work was partially supported by DGI (Spain):
FIS2008-01143, FPA2006-13807-C02-01, FIS2008-04189, FPA2007-62216,
by the Junta de Andaluc\'{\i}a, 
by the INFN-MEC collaboration agreement,
project ``Study of relativistic dynamcis in neutrino and
electron scattering'', the Spanish Consolider-Ingenio
programme CPAN (CSD2007-00042),
and part (TWD) by U.S. Department of Energy under cooperative
agreement DE-FC02-94ER40818.
\end{theacknowledgments}


\begin{thebibliography}{9}

\bibitem{Ama05b} J.E. Amaro, M.B. Barbaro, J.A. Caballero,
T.W. Donnelly, and C. Maieron, Phys. Rev. C 71, 065501 (2005).

\bibitem{Nie04} J. Nieves, J.E. Amaro and M. Valverde.
Phys. Rev. C 70, 055503 (2004).

\bibitem{Spe80} J. Speth, V. Klemt, J. Wambach and G.E. Brown, Nucl. Phys. A343, 382 (1980)

\bibitem{Aue02} L.B. Auerbach {\em et al.,} Phys. Rev. C 66, 015501 (2002).
\bibitem{Ath97} C. Athanassopoulos {\em et al.,} Phys. Rev. C 55, 2078 (1997).

\bibitem{Fer92} P. Fernadez de Cordoba and E. Oset, Phys. Rev. C 46 (1992)1697.

\bibitem{Val06} M. Valverde, J.E. Amaro, J. Nieves, Phys. Lett. B 638 (2006)

\bibitem{Val06b} M. Valverde, J.E. Amaro, J. Nieves, C. Maieron, 
Phys. Lett. B 642 (2006) 218

\bibitem{Mig03} P. Migliozzi, Int. J. Mod. Phys. A 18 (2003) 3877.

\bibitem{Hag03} K. Hagiwara K. Mawatari and H. Yokoya, Nucl. Phys. B 668 (2003) 364. 

\bibitem{Mai02} C. Maieron, T.W. Donnelly and I. Sick, Phys. Rev. C 65
(2002) 025502.

\bibitem{Ama05} J.E. Amaro, M.B. Barbaro, J.A. caballero,
T.W. Donnelly, A. Molinari, and I. Sick, Phys. Rev. C 71, 015501
(2005)


\bibitem{Ama07} J.E. Amaro, M.B. Barbaro, J.A. Caballero,
T.W. Donnelly, and J.M. Udias, Phys. Rev. C 75, 034613 (2007).


\end{thebibliography}
\end{document}